\documentclass[graybox, envcountchap]{svmult}

\usepackage{makeidx}         
\usepackage{graphicx}        
\usepackage{multicol}        
\usepackage[bottom]{footmisc}

\usepackage{newtxtext}       %
\usepackage[varvw]{newtxmath}       

\usepackage{bm, amsmath, latexsym}

\usepackage[acronym]{glossaries}
\makeglossaries
\newacronym{ite}{ITE}{individual treatment effect}
\newacronym{rct}{RCT}{randomized controlled trial}
\newacronym{ate}{ATE}{average treatment effect}
\newacronym{scd}{SCD}{single-case design}
\newacronym{dht}{DHT}{digital health technology}
\newacronym{rite}{RITE}{recurring individual treatment effect}
\newacronym{cate}{CATE}{conditional \acrshort{ate}}
\newacronym{hte}{HTE}{heterogeneous treatment effect}
\newacronym{ich}{ICH}{International Council for Harmonisation}
\newacronym{jitai}{JITAI}{just-in-time adaptive intervention}
\newacronym{mrt}{MRT}{micro-randomized trial}
\newacronym{sced}{SCED}{single-case experimental design}
\newacronym{apte}{APTE}{average period treatment effect}
\newacronym{bg}{BG}{blood glucose}
\newacronym{pte}{PTE}{period treatment effect}
\newacronym{cee}{CEE}{causal excursion effect}
\newacronym{arco}{ARCO}{autoregressive carryover}
\newacronym{motr}{MoTR}{model-twin randomization}
\newacronym{glmm}{GLMM}{generalized linear mixed model}
\newacronym{glm}{GLM}{generalized linear model}
\newacronym{fosr}{FoSR}{function-on-scalar regression}
\newacronym{bhm}{BHM}{Bayesian hierarchical model}
\newacronym{fda}{FDA}{Food and Drug Administration}
\newacronym{smart}{SMART}{sequential multiple assignment randomized trial}
\newacronym{app}{app}{mobile application}

\makeindex             

\begin{document}

\frontmatter

\tableofcontents

%
%
%

%
%
%
%
%
%
%

\title{A Primer on Digital Health N-of-1 Studies and Single-Case Designs}
\author{Eric J. Daza\orcidID{0000-0002-8376-1600}}
\institute{
Eric J. Daza \at Stats-of-1, California, USA, \email{ericjdaza@statsof1.org}
}
%
%
\maketitle

\abstract*{
Clinical studies generally assume that group-level averages are useful quantities for guiding individual-level decisions in the clinical care of each individual patient. Precision medicine has notably closed the gap towards truly individualized care through highly refined subgrouping. Today, digital health technologies and other modern sources of dense, personal ``small data'' enable a different approach to treatment individualization---one that seeks to characterize a single person's own recurring health patterns first and foremost, rather than identifying the best subgroup to which they might belong. In this chapter, we review the key concepts underlying n-of-1 studies, single-case designs, and other ``multitudinal'' approaches for digital health applications, and explore their relationships to other digital health methods. We also share some promising future directions for ``esametry'', the statistics of the digitized multitudes within each of us. 
}

\abstract{
Clinical studies generally assume that group-level averages are useful quantities for guiding individual-level decisions in the clinical care of each individual patient. Precision medicine has notably closed the gap towards truly individualized care through highly refined subgrouping. Today, digital health technologies and other modern sources of dense, personal ``small data'' enable a different approach to treatment individualization---one that seeks to characterize a single person's own recurring health patterns first and foremost, rather than identifying the best subgroup to which they might belong. In this chapter, we review the key concepts underlying n-of-1 studies, single-case designs, and other ``multitudinal'' approaches for digital health applications, and explore their relationships to other digital health methods. We also share some promising future directions for ``esametry'', the statistics of the digitized multitudes within each of us. 
}

\begin{quote}
{\it But what is contemporary technology if not a mechanism for the containment of multitudes?}
\begin{flushright}
--- Taylor Fang \cite{fang2019}
\end{flushright}
\end{quote}


It is not unreasonable to claim that the clinical quantity of most interest is the \acrfull{ite} for each patient: The impact of treatment for each person defined as the difference in their outcomes had they taken one treatment versus another. But only one of these {\sl potential outcomes} \cite{splawa1990application, rubin1974estimating}, the one corresponding to the treatment the patient actually gets, can be observed at any given time! This is the {\sl fundamental problem of causal inference} \cite{holland1986statistics} that drives all of clinical science.

Equipoise compels clinical researchers to find ways to estimate each \acrshort{ite}. The standard approach is to design a \acrfull{rct} or observational study that requires certain conditions to hold under which the \acrshort{ite} does not vary much across study participants. In the ideal case, these study conditions render each \acrshort{ite} exactly equal to the \acrfull{ate} across all participants, a quantity that {\sl can} be estimated.

The \acrshort{ate} is the core statistical estimand around which these nomothetic (i.e., group-focused, group-level) \cite{ponterotto2005qualitative} studies are designed. It is the de facto statistical engine of clinical trials.\footnote{Core parts of clinical science like formative biomedical research, molecule discovery, preclinical studies, pharmacokinetics, pharmacodynamics, and biomarker development don't necessarily focus on groups of patients. However, the ultimate statistical goal of a clinical trial has historically been generalizability (i.e., inference) over a population of patients.} But there is another important way to approach the \acrshort{ite}---one that we believe is more intuitive and philosophically consistent with this uniquely personal quantity.

In this chapter, we will cover how to use n-of-1 studies, \acrlong{scd}s (\acrshort{scd}s), and other ``multitudinal designs'' to analyze data from digital health technologies (\acrshort{dht}s). We provide a brief summary of individualization approaches in Section \ref{sec:rite_quantity}, and an introduction to the causal concepts underlying n-of-1 studies and \acrshort{scd}s in Section \ref{sec:recurring_causal_effects}. Common study designs are briefly reviewed in Section \ref{sec:study_designs_and_models}, along with n-of-1 models and their connection to functional data analysis. We conclude in Section \ref{sec:esametric_clinical_research} by considering how to advance esametric clinical science.


\section{The \acrshort{rite} Quantity}
\label{sec:rite_quantity}


The goal of both n-of-1 studies and \acrshort{scd}s is to characterize idiographic (i.e., individual-focused, individual-level) \cite{ponterotto2005qualitative} treatment effects and statistical associations, quantities specific to each participant. These {\sl recurrence-based approaches} to \acrshort{ite} characterization rely on the existence of a repeatably measurable individual-level effect, and their statistical objective is to estimate a {\sl \acrlong{rite}} (\acrshort{rite}) for each participant \cite{daza2025model}.

A \acrshort{rite} is defined over an {\sl idiographic target population} comprised of ``all possible observation periods wherein the subject is at risk for the chronic health condition, and is under at least one of the treatment conditions being studied''---a sort of ``population of one'' \cite{daza2025model, daza2018causal}. This type of population can accurately be described as {\sl multitudinal} \cite{2024_daza}.\footnote{``Multitudinal'' is an apt descriptor not just by its dictionary definition. It is also a fortunate portmanteau of ``multitude'' (which describes the multiple quantities being discussed) and ``longitudinal'' (which describes the recurring nature of these quantities).} Henceforth, we will use ``multitudinal'' to describe recurrence-based approaches.

\subsection{Top-Down Individualization via Subgrouping}
\label{subsec:top_down}

At baseline, eligibility criteria comprise a set of the aforementioned study conditions. Understood this way, the treatment effect estimated by the study is actually a \acrfull{cate}. This interpretation can help researchers assess how well the \acrshort{ate} will generalize or {\sl transport} to other patient populations with different ``eligiblity conditions''. Subgroup analyses invoke further conditioning, and are essentially ways to assess possible \acrlong{hte}s (\acrshort{hte}s) among the study participants. ``Precision medicine'', which typically tailors treatments that were originally designed to work for groups, is a particularly well-developed modern subgrouping technique.

\bigskip
\noindent
\textbf{The Estimands Framework}
\bigskip

The \acrfull{ich} recently introduced the estimands framework \cite{guideline2017addendum}.\footnote{In this framework, the definition of ``estimand'' is more precise than that of ``statistical estimand''. It requires detailed specification of the treatment, intercurrent events and how to handle them, the intended or target population, outcome variables (e.g., used to define the endpoint), and the population-level summary of the outcome variables. From \acrshort{ich} E9(R1): ``Intercurrent events are events occurring after treatment initiation that affect either the interpretation or the existence of the measurements associated with the clinical question of interest. It is necessary to address intercurrent events when describing the clinical question of interest in order to precisely define the treatment effect that is to be estimated.''}
The \acrshort{ich} E9(R1) guideline proposed strategies for handling post-randomization {\sl intercurrent events} (see footnote) that essentially reflect {\sl decision points} about each individual patient's treatment plan based on their own history of clinical events. For example, if the patient experiences a particular adverse event, then active treatment is to be discontinued and rescue medication should be initiated.

In a sense, these intercurrent-event-handling strategies collectively describe a {\sl retroactive subgrouping} approach. After study data collection, each participant is explicitly assigned to the subgroup of participants with a particular intercurrent event that was defined a priori.\footnote{These predefined subgroups can be thought to index potential outcomes; e.g., one potential outcome when treatment continues throughout the entire trial, and a distinct potential outcome when treatment is followed by rescue medication.} The \acrshort{ate} is then estimated over all subgroups based on randomized treatment at baseline per the intention-to-treat (ITT) principle.

\bigskip
\noindent
\textbf{Tailoring the Treatment}
\bigskip

Compare this with the {\sl proactive subgrouping} of a treatment-tailoring approach like a dynamic treatment regime \cite{murphy2003optimal} or \acrfull{jitai} \cite{nahum2016just}. These adaptive treatment strategies are comprised of a treatment regime (i.e., set of treatment rules) used to tailor the intervention for a given individual at a given decision point based on their relevant health history and availability at that point in time \cite{zhang2019near}. The main interest lies in characterizing how well a given ``treatment rulebook'' can improve health outcomes on average. Contrast this with the common \acrshort{rct} goal of characterizing the average effect of a set treatment.

A \acrfull{mrt} \cite{walton2018optimizing} is used to answer the more granular scientific question, ``For a group of individuals, what is the average effect of treatment $A$ on clinical outcome $Y$ at any given decision point after randomizing $A$ at every previous decision point according to a given treatment regime?'' In an \acrshort{mrt}, our main goal is to estimate an average effect across all study participants at a given clinical decision point. The causal estimand of an \acrshort{mrt} is the result of an active, sequential process for individualizing the intervention.

\subsection{Bottom-Up Individualization via Recurrence}
\label{subsec:bottom_up}

If the {\sl subgrouping-based approaches} above take a ``top-down'' approach to defining an estimable type of \acrshort{ite}, then an n-of-1 study or \acrshort{scd} takes a ``bottom-up'' approach \cite{daza2025model, vegetabile2021distinction}. An n-of-1 study seeks to answer a fundamentally different question: ``For a {\sl particular individual}, what is the average effect of $A$ on clinical outcome $Y$ at any given treatment period if we were to randomize $A$ at every treatment period?'' An \acrshort{scd} asks a similar question, ``For a particular individual, what is the average effect of intervention $A$ on clinical outcome $Y$ at any given {\sl phase} if we were to administer either intervention or no intervention over different phases?''

\bigskip
\noindent
\textbf{Multitudinal Designs}
\bigskip

N-of-1 studies and \acrshort{scd}s are {\sl multitudinal designs} used in biomedical clinical research, in clinical care for dose-finding or dose-tailoring, and in clinical, educational, and behavioral psychology \cite{scuffham2010using, gabler2011n, smith2012single, mirza2017history, shaffer2018n, selker2022useful, Selker2023, 2023_defelippe_etal}. These study designs have also been gaining traction in patient-centered clinical trials, buoyed by two definitive design guides \cite{kravitz2014design, nikles2015essential} and a CONSORT extension \cite{Vohra2015}.

In a typical n-of-1 trial, each participant is experimentally assigned a treatment over multiple treatment periods. Only one outcome measurement per period is typically used in analysis. Selker et al (2022, 2023) call these {\sl Type 1} n-of-1 trials \cite{selker2022useful, Selker2023};  we will use ``n-of-1 study'' and ``Type 1 n-of-1 study'' interchangeably. A typical \acrfull{sced} includes two to four treatment phases, with multiple outcome measurements per phase used in analysis \cite{kratochwill2013single, epstein2022family}. Selker et al conversely call these {\sl Type 2} n-of-1 trials; we will interchangeably use ``single-case design'' and ``Type 2 n-of-1 study''. For both Types of n-of-1 trial, a treatment level is allowed to {\sl cross over} to a different level between periods or phases.

In biomedicine and clinical settings, n-of-1 trials are commonly used for the diagnosis, treatment, management, or prevention of idiosyncratic chronic diseases for which an \acrshort{ate} is less clinically meaningful or useful because of extreme effect heterogeneity \cite{davidson2022introducing}. These are sometimes called {\sl reversal designs} in psychology and education because the treatment effect can {\sl wash out}, reverse, or otherwise return to baseline before a new treatment level is then assigned \cite{epstein2022family}. N-of-1 trials are called by names such as {\sl switchback experiments} or {\sl within-subject experimental designs} in other fields like finance, business, economics, and ecology \cite{eichler2012causal, moraffah2021causal, runge2023causal, bojinov2023design, list2025experimentalist}.

In biomedical research, \acrshort{sced}s are used to understand and treat rare diseases. Treatment crossover may only happen once; e.g., a baseline or non-intervention phase followed by an intervention phase \cite{linden2018matching, muller2021systematic}. This infrequent crossover structure is also the main design used in psychology and education, where reversal of the treatment effect is often not possible. For example, it may be difficult to unlearn or otherwise forget a skill taught by an educational intervention.

A Type 2 n-of-1 trial for patients with rare diseases can also resemble \acrshort{jitai}. In particular, consider how combination therapies, antisense oligonucleotides (ASOs), and other precision medicine techniques can be used at baseline to tailor the treatment to each patient based on their unique genetic variants \cite{sicklick2021molecular, ctgov20260512eommp1aso}. The main question of interest is how well the tailored treatments work on average across all participants.

Common design and inference challenges in n-of-1 studies and \acrshort{scd}s include {\sl autocorrelation}, treatment effect {\sl carryover} from past periods, and how to implement blinding in a single participant \cite{kravitz2014design, nikles2015essential}. Study data are structured as multivariate, multilevel time series \cite{goldstein1994multilevel} with possible non-stationarity over time \cite{cai2024causal, fowler2024testing, fowler2025individual}.

\bigskip
\noindent
\textbf{Aggregating Individual Findings}
\bigskip

It is often easier to understand a new idea or approach that is explained using well-established concepts. In a multitudinal design, the repeated measurements comprise the study sample. The single study participant would be the ultimate ``subgroup'' of a multi-participant study. Can we somehow combine these single-person studies to draw more general conclusions about groups of participants? Is there a way to bridge idiographic and nomothetic study designs?

Indeed, a {\sl series of n-of-1 trials} is often aggregated to conduct population-level inference, often via meta-analysis or hierarchical modeling \cite{1997_zucker_etal, 2010_zucker_etal, Kratochwill2014}. Think of how fitting separate individual-specific n-of-1 trial models reflects a ``no-pooling'' approach in terms of mixed-effects models. Then one way to aggregate findings across n-of-1 trial participants is through ``partial pooling'' of information: a common population-level model structure is imposed across all participants, letting only the random terms vary across individuals.

But in \textit{The End of Average} \cite{rose2016end}, Todd Rose rightly cautioned that in many important cases, the average ``creates the illusion of knowledge, when in fact [it] disguises what is most important about an individual.'' How might aggregation affect the quality of individual-level inference?

The {\sl ergodicity information index} is one way to assess whether to use a multitudinal or subgrouping-based design \cite{golino2025toward, kaiser2025must}. It ``quantifies the amount of information lost by representing all individuals with a between-person structure''. More flexible aggregation techniques may better preserve idiographic findings, such as ones that use generalized additive linear mixed models \cite{grekov2025flexible}. And Bayesian approaches can be used to incorporate individual-specific prior information into a mixed-effects model \cite{2010_zucker_etal}.

\bigskip
\noindent
\textbf{Esametry: The Statistics of One}
\bigskip

N-of-1 studies, \acrshort{scd}s, and other multitudinal designs are the foundation of the emerging area of statistics that we call {\sl esametry}\footnote{From Daza et al (2025): The term is ``derived from `isa' (pronounced `ee-SA'), the Tagalog Filipino word for `one'.'' Esametry is an English word pronounced like ``ee-SA-met-tree''.}, defined as ``the application of statistics to a single person, individual, or unit''. Esametry is the set of broadly defined {\sl quantitative idiographic} \cite{shoda2013cognitive} approaches applied to digital health technologies and other modern sources of {\sl intensive longitudinal data} \cite{2006_walls_schafer}. In addition to the temporal challenges mentioned earlier and in Section \ref{subsec:temporal_considerations} below, modern esametric studies must often contend with the extensive and complex missing data patterns found in dense longitudinal data from \acrshort{dht}s \cite{cai2025missing}.

\section{Recurring Causal Effects}
\label{sec:recurring_causal_effects}

The fundamental data structure of a multitudinal study is a multivariate time series. Estimation and inference must often account for temporal phenomena like autocorrelation within variables and serial cross-correlation between variables. Some of these complications may be mitigated or removed through a well-designed n-of-1 trial or \acrshort{sced}. However, digital health data are often collected in non-randomized, ecological, or real-world settings. Hence, it will help to understand the relevant causal quantities.

A multitudinal study focuses on analyzing a single participant measured repeatedly over periods $t = 1, \dots, m$ (for $m>1$).\footnote{One can think of a multitudinal study as an ``n-of-1, m-of-many''.} In this section, we define the \acrshort{rite} estimand for n-of-1 studies and propose a similar estimand for \acrshort{scd}s. We will also discover a core concept shared by \acrshort{mrt}s.

We will occasionally refer to related concepts from other chapters like those on \acrshort{mrt}s and functional data analysis. To aid cross-conceptual understanding, we have tried to mirror some of their notation throughout this chapter.

Note that formal causal inference for multitudinal studies is still early in development. There is currently no consensus on terminology and notation. Herein, we present our understanding of multitudinal causal inference based on the {\sl \acrlong{apte}} (\acrshort{apte}) framework of Daza (2018) \cite{daza2018causal} and Daza et al (2025) \cite{daza2025model}, defined in Section \ref{subsec:nof1_studies}. But for esametry to mature, these germinal concepts must be challenged, tested, and refined.

The reader should therefore explore other approaches and frameworks, in particular Bojinov and Shephard (2019) \cite{bojinov2019time} on time series experiments applied to financial trading, Anjum et al (2020) \cite{anjum2020rethinking} on the philosophy of causation for each individual patient, Malenica et al (2021, 2023) \cite{malenica2021adaptive, malenica2023anytime} on time series adaptive sequential designs and n-of-1 trial anytime-valid inference, the U-\acrshort{cate} framework of Piccininni et al (2024) \cite{piccininni2024causal}, Kaiser et al (2026) \cite{kaiser2026attributing} on individual-level causal effects in education and psychology, and many other references mentioned throughout this chapter.

\subsection{Running Example: SleepBG}
\label{subsec:running_example}

Daza et al (2019) \cite{daza2019effects} conducted n-of-1 trials on two of their authors. Each n-of-1 trial consisted of eight treatment periods, with each period comprised of three {\sl sub-period} days. Henceforth, we will refer to this as the SleepBG study.

Their primary clinical objective was to estimate the \acrshort{apte} of sleep deprivation on continuous \acrfull{bg} level, mood, and food cravings in the two non-diabetic study participants. Sleep deprivation was defined and enforced as no more than four hours of overnight sleep during the {\sl treatment sub-period}, followed by at least six hours of overnight sleep to recover during the two subsequent {\sl washout sub-periods}.\footnote{See Section \ref{subsec:temporal_considerations}.} The baseline or control condition was getting at least six hours of sleep on all three nights.

Study participants used a continuous glucose monitoring device to record their \acrshort{bg} levels (mg/dL). Their mood was assessed using the five-point Positive and Negative Affect Schedule (PANAS) scale \cite{watson1988development} twice a day (i.e., beginning and end). Craving, defined as ``an intense desire to consume a particular food or food type that is difficult to resist'', was assessed three times a day (i.e., beginning, middle, and end).

Henceforth, we will only reference the analyses involving \acrshort{bg} levels, and only consider the treatment sub-period at each period $t$ unless otherwise noted. The authors used log-transformed \acrshort{bg} (log-\acrshort{bg}) to better meet the normality assumptions used in their modeling. Only one of the authors provided enough study data for analysis.

\subsection{N-of-1 Studies}
\label{subsec:nof1_studies}

The outcome analyzed in SleepBG was the log-\acrshort{bg} level at time point $s$ over the three study days of period $t+1$ for all $m = 8$ periods, which we denote as $Y_{t+1} ( s )$. Log-\acrshort{bg} was measured every 14 minutes on average (i.e., median), corresponding to about 103 time points on any given study day. The binary treatment was sleep deprivation the night before, denoted $A_t = 1$, or its baseline counterpart, denoted $A_t = 0$.

In this section, we consider the simpler case with only one time point per period, and we write $Y_{t+1}$ to represent the outcome of interest. We will return to the temporally dense case with multiple time points per period in Section \ref{subsec:functional_data_analysis} (on functional data analysis).

Let $\bm{V}_t$ represent a vector of recurring time-varying covariates consisting of the following two mutually exclusive sets of variables. Let $\bm{V}^\text{ex}_t$ denote exogenous variables that affect (cause) $A$ or $Y$, but are never themselves affected by $A$ or $Y$; for example, ambient temperature. Conversely, let $\bm{V}^\text{en}_t$ denote endogenous variables other than the outcome; for example, suppose $V_t$ affects $Y_{t+1}$, which in turn affects $V_{t+1}$, and so on. At any period $t$, define the temporal order of events from earliest to latest as $\bm{V}_t, A_t, Y_{t+1}$.

Let $\bm{H}_t = ( \bm{V}_1, A_1, Y_2, \dots, \bm{V}_t, A_{t-1}, Y_t )$ represent the {\sl history prior to $A_t$}. Let $\overline{Y}_t = ( Y_1, \dots, Y_t )$ denote all possible lagged outcomes that would be part of the history $\bm{H}_t$. Define $\overline{A}_{t-1}$ and $\overline{\bm{V}}_t$ likewise.

\bigskip
\noindent
\textbf{Average Period Treatment Effect}
\bigskip

The potential outcome at period $t+1$ that would be observed under treatment level $A_t = a_t$ with history $\bm{H}_t$ is denoted $Y_{t+1} ( a_t )$. In the n-of-1 setting, the \acrshort{ite} is defined at each period rather than for each participant. Specifically, the \acrshort{ite} is defined as a difference between $Y_{t+1} (1)$ and $Y_{t+1} (0)$. We will henceforth call this quantity the {\sl \acrlong{pte}} (\acrshort{pte}). The \acrshort{pte} on a given study day in SleepBG was the difference in log-\acrshort{bg} between the two possible conditions wherein the study participant did and did not experience sleep deprivation the night before.

In an n-of-1 trial, treatment is randomized at every period, and the \acrshort{pte} we will consider is the mean difference
\begin{align*}
\delta^\text{\acrshort{pte}}_{t+1}
    &= E_{\overline{A}_{t-1}} \left\{
            Y_{t+1} \left( \overline{A}_{t-1}, 1 \right) -
            Y_{t+1} \left( \overline{A}_{t-1}, 0 \right)
            \Big|
            \overline{\bm{V}}^\text{ex}_t
        \right\}
.
\end{align*}
The \acrshort{rite} estimated by n-of-1 studies is called the average period treatment effect, a within-individual average quantity defined as the average \acrshort{pte} taken across the full study period.\footnote{See Appendix Section \ref{subsec:more_on_causal_estimands} for the explicit \acrshort{apte} formula.} Note that {\sl the \acrshort{apte} is not an \acrshort{ate}}; it is an average taken over periods for one individual, while the \acrshort{ate} is an average taken over individuals.\footnote{Daza (2018) coined the term ``average {\sl period} treatment effect'' precisely for this reason; i.e., to distinguish it from the ``average treatment effect'', which is commonly defined over a group of people.}

In SleepBG, sleep deprivation was initially estimated to increase the participant's log-\acrshort{bg} level by $0.011$ on average at any given period, based on a priori assumptions $(p = 0.0230)$. However, post hoc adjustments to the model revealed that variations in \acrshort{bg} level were most explained by predictors other than treatment; there was no statistically discernible\footnote{There was not enough statistical evidence to ``discern'' a true non-zero \acrshort{apte}; i.e., the p-value of the estimated \acrshort{apte} was not statistically significant \cite{daza2019effects, barnett1997tyranny, kuhberger2015significance, amrhein2019scientists, mcshane2019abandon, 2019_witmer, matias2022possibly}.} \acrshort{apte} of sleep deprivation in the updated model $(p = 0.66)$.

Importantly, the \acrshort{pte} formula explicitly states the conditions under which both the \acrshort{pte} and \acrshort{apte} are generalizable. They transport to settings wherein the exogenous covariates at any period $( \bm{V}^\text{ex}_t )$ are similarly distributed to those observed in the study; for example, under similar longitudinal patterns of ambient temperature as those observed during the full study period.

\bigskip
\noindent
\textbf{Causal Excursion Effect vs. Period Treatment Effect}
\bigskip

The \acrshort{pte} above resembles the \acrfull{cee} of an \acrshort{mrt} \cite{boruvka2018assessing} when the participant is always available for treatment assignment. Where the two quantities diverge is in their assumptions, which reflect the distinct scientific goals of \acrshort{mrt}s and multitudinal studies.

At a clinical decision point in an \acrshort{mrt}, each participant is implicitly assigned to the subgroup created by their own health history over all past decision points; for example, by their sequence of intercurrent events.\footnote{Recalling the \acrshort{rct} design, the set of intercurrent-event-handling strategies can loosely be thought of as a retrospectively determined non-randomized \acrshort{jitai} treatment regime over irregularly spaced clinical decision points.} The \acrshort{jitai} tailoring process, like the \acrshort{rct} estimands framework, also serves to estimate an overall average effect across individuals. At a given decision point, the outcomes can be averaged per treatment arm and then compared, yielding an estimate of the overall or {\sl marginal \acrshort{cee}}. Estimating this ``tailored \acrshort{ate}'' is the goal of an \acrshort{mrt}.

The \acrshort{cee} within a given subgroup (i.e.,  conditional on a particular health history) is the HTE for that subgroup. This {\sl conditional \acrshort{cee}} can be thought of as a subgroup-specific average \acrshort{ite} shared among that subgroup's members. Although reducing variability across sequential subgroups can be desirable (at least for inference, if not clinically), a \acrshort{jitai} in general does not require these conditional \acrshort{cee}s to converge to the marginal \acrshort{cee} (or to converge at all).

We are often interested in finding a marginal \acrshort{cee} that is more clinically meaningful or less variable than the \acrshort{ate} that would have been observed without tailoring. Or perhaps patients are more likely to engage with and adhere to the treatment regime because of its personalization components, which can help reduce bias and variance in effect estimation.

But we aren't particularly interested in characterizing a \acrshort{cee} that doesn't change from period to period. A \acrshort{jitai} does not necessarily create a constant or stable individual-specific treatment effect that can be repeatedly measured.\footnote{That said, a \acrshort{jitai} with a long-enough duration may indeed create stable, recurring ITEs. A treatment regime over a long study duration can create a large number of subgroups, each with only a few study participants. For example, suppose the effects vary little among members of a subgroup with a shared health history. If these effects remains fairly stable over multiple repeated randomizations after a certain decision point, then their average can be thought of as a stable conditional \acrshort{cee}. The set of decision points over which such a stable effect exists can partially describe the multitudinal target population defined in Section \ref{sec:rite_quantity}.} If SleepBG had been a \acrshort{jitai} that tailored the amount of sleep deprivation to assign each night based on each participant's \acrshort{bg} levels the day before, the investigators may have wanted to estimate a \acrshort{cee} rather than a \acrshort{pte}.

In contrast to an \acrshort{mrt}, the goal of a multitudinal study is to estimate a recurring average effect across all study time periods for one person. The \acrshort{pte} is an average quantity taken over all prior treatment paths for only one study participant. And the resulting \acrshort{apte} is assumed to be a constant, stable, recurring quantity---at least during the study, if not in the long run.\footnote{See Appendix Section \ref{subsec:estimable_aptes} for a definition of ``stable'' in this context.}

\subsection{Single-Case Designs}
\label{subsec:scd_causal_estimand}

The fundamental data structure of a \acrshort{scd} is an interrupted time series \cite{kratochwill2010single}. As mentioned in Section \ref{subsec:bottom_up}, the basic design involves observing the repeated outcome during a baseline non-intervention phase, administering an active treatment, then observing the repeated outcome during the subsequent intervention phase.

In \acrshort{scd}s, the preferred term for ``period'' is ``phase''. To avoid terminological confusion in this section, we will use $t$ to index a time point rather than a period. Define a {\sl phase} as a contiguous set of time points where we set $a_t = 0$ during a baseline phase and $a_t = 1$ during an intervention phase.

\bigskip
\noindent
\textbf{Reversal Designs}
\bigskip

The most basic \acrshort{scd} consists of one crossover. The baseline phase is typically denoted ``A'', and it often precedes the intervention phase, typically denoted ``B'', so this is frequently called an AB design. For example, suppose the baseline phase has two time points, followed by an intervention phase with four time points. Then the treatment vector is defined over six time points as $\overline{a}_6 = ( 0, 0, 1, 1, 1, 1 )$. The reversal design extends the AB design by alternating the baseline and intervention phases multiple times.

For example, Lopes et al (2023) \cite{lopes2023exploring} published a study protocol with an ABAB design, which has three crossovers and the corresponding treatment vector $\overline{a}_4 = ( 0, 1, 0, 1 )$. Their objective is ``to explore the efficacy of a set of smart devices to detect malposture and increase postural awareness, reducing fatigue, and musculoskeletal disorders'' in five industrial manufacturing workers in Portugal. During the two intervention ``B'' phases, each worker will use a smart wearable system comprised of a jacket, footwear, band, and feedback system.

The study's primary outcomes will be fatigue assessed via electromyography, and musculoskeletal symptoms characterized using the Nordic Musculoskeletal Questionnaire \cite{mesquita2010portuguese}. Secondary outcomes will be assessed via motion analysis \cite{dahl2020wearable}, structured visual analysis \cite{bulte2013single}, clinical assessment \cite{mcatamney1993rula}, and other patient-reported outcomes (PROs) (i.e., psychometric instruments or questionnaires).

\bigskip
\noindent
\textbf{Multiple Baseline Designs}
\bigskip

A {\sl multiple baseline design} enrolls study participants that start the intervention phase at different times \cite{kazdin1982single, kazdin2019single}. The crossover points vary across the participants. For example, over six time points, one participant might be assigned the example AB treatment vector shown earlier, while another might be assigned $\overline{a}_6 = ( 0, 0, 0, 1, 1, 1 )$. In pharmaceutical drug development, this design might conceptually resemble a dose-finding study wherein each unique intervention-phase duration corresponds to a ``dose level'' \cite{2025_hall_daza}.

Lancioni et al (2025) \cite{lancioni2025technology} combined the reversal and multiple baseline designs to assess the feasibility and effectiveness of using a technology-enabled environmental stimulus delivery system to improve developmental behaviors in eight adult participants with severe intellectual disabilities and sensorimotor impairments. The authors used an ABACB reversal design wherein each phase consisted of multiple session blocks (i.e., time points), with each session lasting five minutes. This was also a multiple baseline design because the length of each phase varied across participants. Before the ABACB phases, the study team assessed each participant to select which of three 10-second audio-visual segments they preferred as environmental stimulation.

At each session during the two intervention ``B'' phases, a technology system comprised of a ``webcam sensor linked to a portable computer, a Bluetooth mini speaker, and a smart Wi-Fi plug'' provided a participant with 10 seconds of their preferred stimulation whenever the system recorded a response. A control ``C'' phase was included to assess the degree to which an intervention effect was due to stimulation itself rather than the response-adaptive stimulation structure of the intervention. This was done by providing a participant with stimulation (preferred or otherwise) throughout each control session regardless of their response pattern.

Figures 2 and 3 of Lancioni et al (2025) illustrate how the intervention produced more frequent responses across participants when compared against baseline and control conditions, with a possible practice effect from habituation observed in the last intervention phase. These figures also showcase the variety of shapes (line plots) across participants of effects over time---not just average differences between baseline/control and intervention conditions.

\bigskip
\noindent
\textbf{Average Period Treatment Effect for Single-Case Designs}
\bigskip

Subsets of phases are often used in \acrshort{scd}s because the intervention may not reach its full effect immediately after being started. This is called {\sl slow onset} in the n-of-1 trial literature \cite{daza2018causal, kravitz2014design, nikles2015essential}. For example, in the second baseline phase of Lopes et al (2023), study outcomes were assessed only after four weeks of washout.\footnote{See Section \ref{subsec:temporal_considerations}.}

Hence, a reasonable \acrshort{scd} \acrshort{apte} might be defined as a difference in average outcomes over stable subsets of the baseline and intervention phases after a slow-onset sub-phase.\footnote{See Appendix Section \ref{subsec:estimable_aptes} for a definition of ``stable'' in this context.} The reader should compare this definition with corresponding quantities defined and derived by Valente et al (2023) \cite{valente2023causal}, who used potential outcomes to refine the statistical theory of causal mediation analysis (which characterizes both direct and indirect treatment effects).

\subsection{Temporal Challenges}
\label{subsec:temporal_considerations}

Autocorrelation exists when a variable at one period is correlated with itself at a different period. In this chapter, we will specifically use ``autocorrelation'' to refer to serial correlation in the outcome. For example, suppose the {\sl data-generating model}\footnote{We use this term as a synonym for {\sl structural causal mechanism} as defined in Appendix Section \ref{subsec:more_on_causal_estimands}.} is $Y_{t+1} = \beta_0 + \beta_\text{A} A_t + \beta_\text{ar} Y_t + \varepsilon_t$, where $\varepsilon_t$ represents completely random error (e.g., due to random between-measurement variation) with zero mean and finite variance. Then autocorrelation exists when the autoregressive coefficient $\beta_\text{ar}$ is not equal to zero.

We say that a time trend exists if the mean outcome increases or decreases across periods. For example, suppose the data-generating model is $Y_{t+1} = \beta_0 + \beta_\text{A} A_t + \beta_\text{tr} t + \varepsilon_t$ with time trend coefficient $\beta_\text{tr} > 0$. Then the mean outcome increases by $\beta_\text{tr}$ with every period.

The U.S. Agency for Healthcare Research and Quality (AHRQ) user's guide on n-of-1 trials defines {\sl treatment carryover} as ``the tendency for treatment effects to linger beyond the crossover (when one treatment is stopped and the next one started)'' \cite{kravitz2014design}. Hence, we refer to the existence of non-zero serial cross-correlation between $A$ and $Y$ (apart from the effect of $A_t$ on $Y_{t+1}$) as a {\sl carryover influence} or simply ``carryover''. If carryover also modifies the \acrshort{pte}, then we say that a {\sl carryover effect} exists.

For example, suppose the data-generating model is $Y_{t+1} = \beta_0 + \beta_\text{A} A_t + \beta_\text{co} A_{t-1} + \beta_\text{Aco} A_t A_{t-1} + \varepsilon_t$. Carryover exists whenever there is a non-zero carryover coefficient $\beta_\text{co}$ or non-zero interaction term $\beta_\text{Aco}$.\footnote{In this example, a causal inference phenomenon called {\sl temporal interference} or {\sl serial interference} exists \cite{daza2025model, wang2021causal, liang2025randomization} because there are more than two potential outcomes at a given period; i.e., $Y_{t+1} ( a_{t-1}, a_t ) \ne Y_{t+1} ( a_{t-1}', a_t )$, where $a_{t-1}' \ne a_{t-1}$. For example, $Y_{t+1} ( 1, 1 ) \ne Y_{t+1} ( 0, 1 )$.} But a carryover effect only exists when $\beta_\text{Aco} \ne 0$. Specifically, if $A_t$ is randomized to 0 or 1 with equal probability, then the \acrshort{pte} is $\delta^\text{\acrshort{pte}}_{t+1} = \beta_\text{A} + \beta_\text{Aco} 0.5$ as shown in Table 1 of Daza et al (2025). That is, carryover modifies the \acrshort{pte} only when $\beta_\text{Aco} \ne 0$.\footnote{In this example, the \acrshort{pte} is constant at every period. This property is called {\sl effect constancy}, which may not hold in general. However, if the \acrshort{pte} is constant in the long run, then we can estimate the \acrshort{apte}. See Appendix Section \ref{subsec:estimable_aptes} for a definition of ``stable'' in this context.}

If the treatment effect duration is known, {\sl washout periods} can be included to reduce or remove carryover effects. Specifically, a washout period is added before the next period to ensure that any intervention effect dissipates and the outcome returns to its baseline level before the next treatment level is assigned. Recall from Sections \ref{subsec:running_example} and \ref{subsec:scd_causal_estimand} that both SleepBG and Lopes et al (2023) used washout periods or sub-periods.

\section{Study Designs and Statistical Considerations}
\label{sec:study_designs_and_models}

A number of common multitudinal study designs account for the aforementioned temporal challenges. These include designs already mentioned such as the n-of-1 trial, phase-change designs (e.g., AB, ABAB/reversal), and the multiple baseline design. Other standard designs include visual analysis, the changing criterion design (which resembles a \acrshort{jitai} study with a single participant), nonoverlap techniques, time series methods like auto-regressive integrated moving average (ARIMA) models, and causal mediation analysis, along with a number of Bayesian approaches \cite{kratochwill2010single, ferron2006tests, Senarathne2020, shrestha2021bayesian, miovcevic2022causal, schmid2022bayesian, 2026_de_carvalho_etal}.

Comprehensive lists of resources and references on esametric methods and study designs can be found online. These include the websites for Stats-of-1 and the International Collaborative Network for N-of-1 Trials and Single-Case Designs (ICN).\footnote{Stats-of-1 (statsof1.org/resources), ICN (nof1sced.org/resources)} Both websites provide numerous resources and references for other statistical considerations such as power and sample size, missing data, blinding, randomization compliance, simulations, and aggregation of multitudinal study findings.

\subsection{N-of-1 Statistical Models}
\label{subsec:recurrence_based_models}

One particularly well-developed model for n-of-1 studies is the dynamic regression model of Schmid (2001) \cite{schmid2001marginal}. This model is based on the time series autoregressive model, and it can be considered a ``Granger model'' because of its resemblance to models used to test for Granger causality \cite{daza2025model}.\footnote{In causal inference terms, ``Granger causality'' is a statistical cross-correlation between two time series that does not necessarily describe an average treatment effect.} The dynamic regression model was extended by Vieira et al (2017) \cite{vieira2017dynamic} to handle binary outcomes, and by Daza et al (2025) as the \acrfull{arco} model that accounts for effect modification due to carryover or autocorrelation.

Daza et al applied an observational causal inference digital twin method called \acrfull{motr} to estimate the \acrshort{apte} of walking cadence (i.e., fast vs slow steps per minute) on sleep duration for two of the authors using their own non-randomized wearable sensor data. For each author, they first fit an \acrshort{arco} model. They then applied \acrshort{motr}: they emulated a target n-of-1 trial by randomizing walking cadence across all observed periods, and then simulating sleep duration at each period by generating the expected sleep duration using the fit \acrshort{arco} model and adding noise to reflect statistical uncertainty.\footnote{The \acrshort{motr} (``motor'') approach is actually quite flexible because it is model-agnostic. The outcome model can take any form that can be stated as a conditional expectation. This includes many non-linear classification and prediction functions used in machine learning and artificial intelligence.} Matias et al (2022) \cite{matias2022possibly} likewise used \acrshort{motr} to estimate heterogeneous \acrshort{apte}s of physical activity on nighttime heart rate based on their own real-world observational \acrshort{dht} data.

These models can be understood as special cases of a {\sl multitudinal \acrlong{glmm}} (\acrshort{glmm}) that allows multiple n-of-1 studies to be aggregated, as was done by the researchers mentioned in Section \ref{subsec:bottom_up} and many others.\footnote{See Appendix Section \ref{subsec:arcoglmm} for the full sketch of a multitudinal \acrshort{glmm}.} A single n-of-1 study would be a special case of a multitudinal \acrshort{glmm}. For example, the data-generating model used to illustrate autocorrelation in Section \ref{subsec:temporal_considerations} is an \acrshort{arco} \acrfull{glm} for continuous outcomes with the identity link function. And Vieira et al (2017) used an \acrshort{arco} \acrshort{glm} with the logit link to predict the putative effects of predictors on the probability of performing a bout of physical activity (where bouts derived from accelerometry data).

\subsection{Multitudinal Functional Data Analysis}
\label{subsec:functional_data_analysis}

Functional data analysis is a statistical technique that is particularly well-suited to modeling a single individual's intensive longitudinal data from \acrshort{dht}s like wearable sensors \cite{acar2025functional}. As with other analysis methods, the standard nomothetic concepts and theory of functional data analysis can be adapted to the recurrent phenomena of idiographic studies. To illustrate, we provide a preliminary sketch of a common functional data analysis model in multitudinal terms.

For a single individual in an n-of-1 study, let $W_t ( s )$ denote the outcome curve (i.e., functional response) at time point $s$ of period $t$. The corresponding data-generating model can be written as the following multitudinal \acrfull{fosr} model:
\begin{align*}
W_{t+1} ( s )
    &= \beta_0 ( s ) + A_t \gamma_A ( s ) + \bm{H}_t \bm{\gamma} ( s ) + \upsilon_t ( s ) + \varepsilon_t ( s )
\end{align*}

Here, $\beta_0 ( s )$ represents the population-level functional intercept, where ``population-level'' means ``across all possible periods'' as defined in Section \ref{subsec:top_down}. The $\gamma ( s )$ terms denote the functional coefficients of the scalar predictors. The history $\bm{H}_t$ may include lagged outcomes contained in $\overline{W}_t$. The outcome curve at a given period can systematically deviate from the average cross-period population-level curve, denoted by the smooth process $\upsilon_t ( s )$. Lastly, $\varepsilon_t ( s )$ represents a smooth error function with zero mean and finite variance.

\begin{figure}[b]
\sidecaption
\includegraphics[scale=0.55]{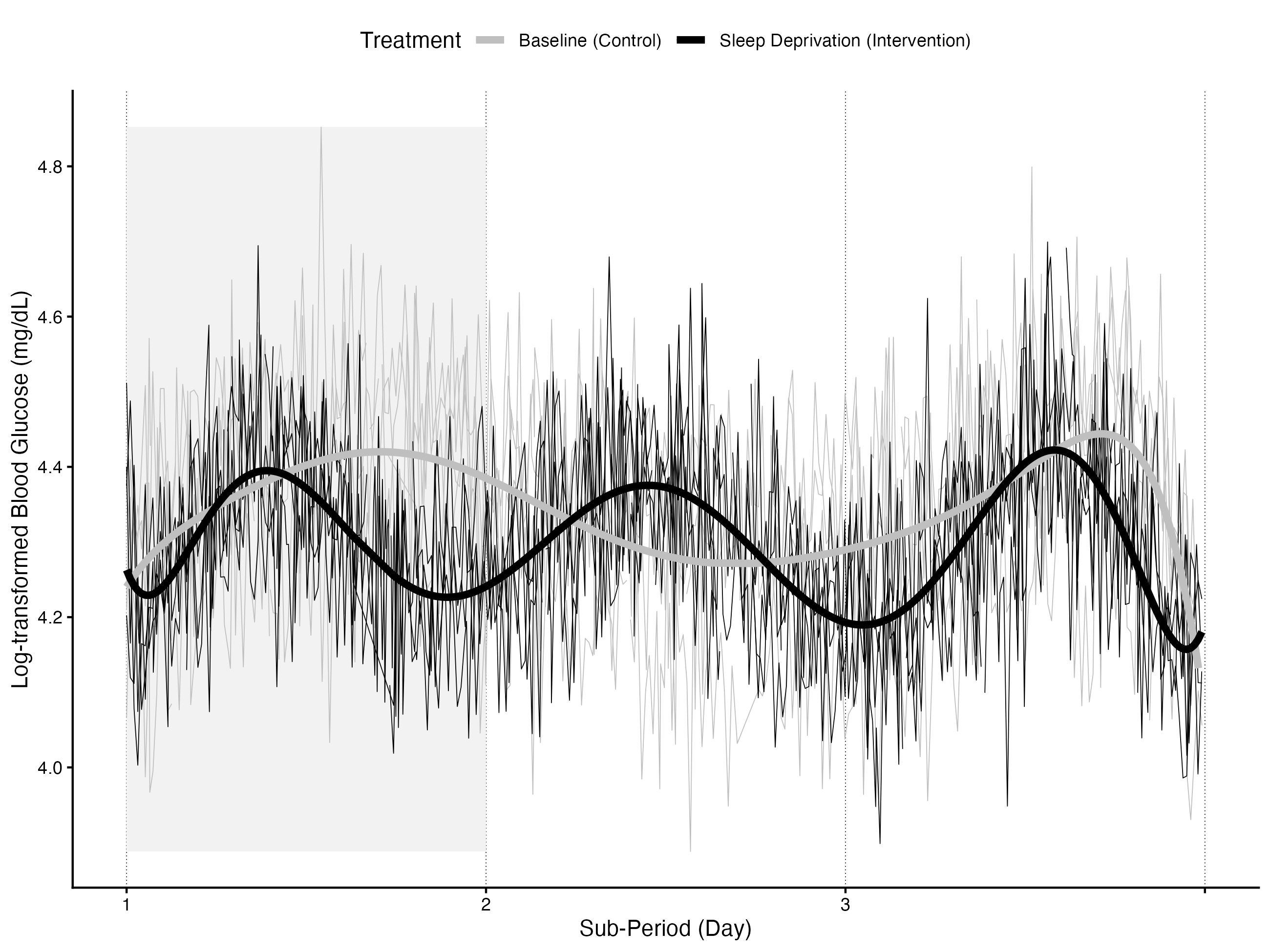}
%
%
\caption{Pancit plot of simulated log-transformed blood glucose with m = 8 treatment periods, each with 1 treatment day (highlighted sub-period) followed by 2 washout days. Thick lines: latent mean curve estimated using a mixed-effects model in Daza et al (2019). Thin lines: observed values simulated around mean curve. Treatment levels: baseline or sleep deprivation.}
\label{fig:pancit}       
\end{figure}

\begin{figure}[b]
\sidecaption
\includegraphics[scale=0.55]{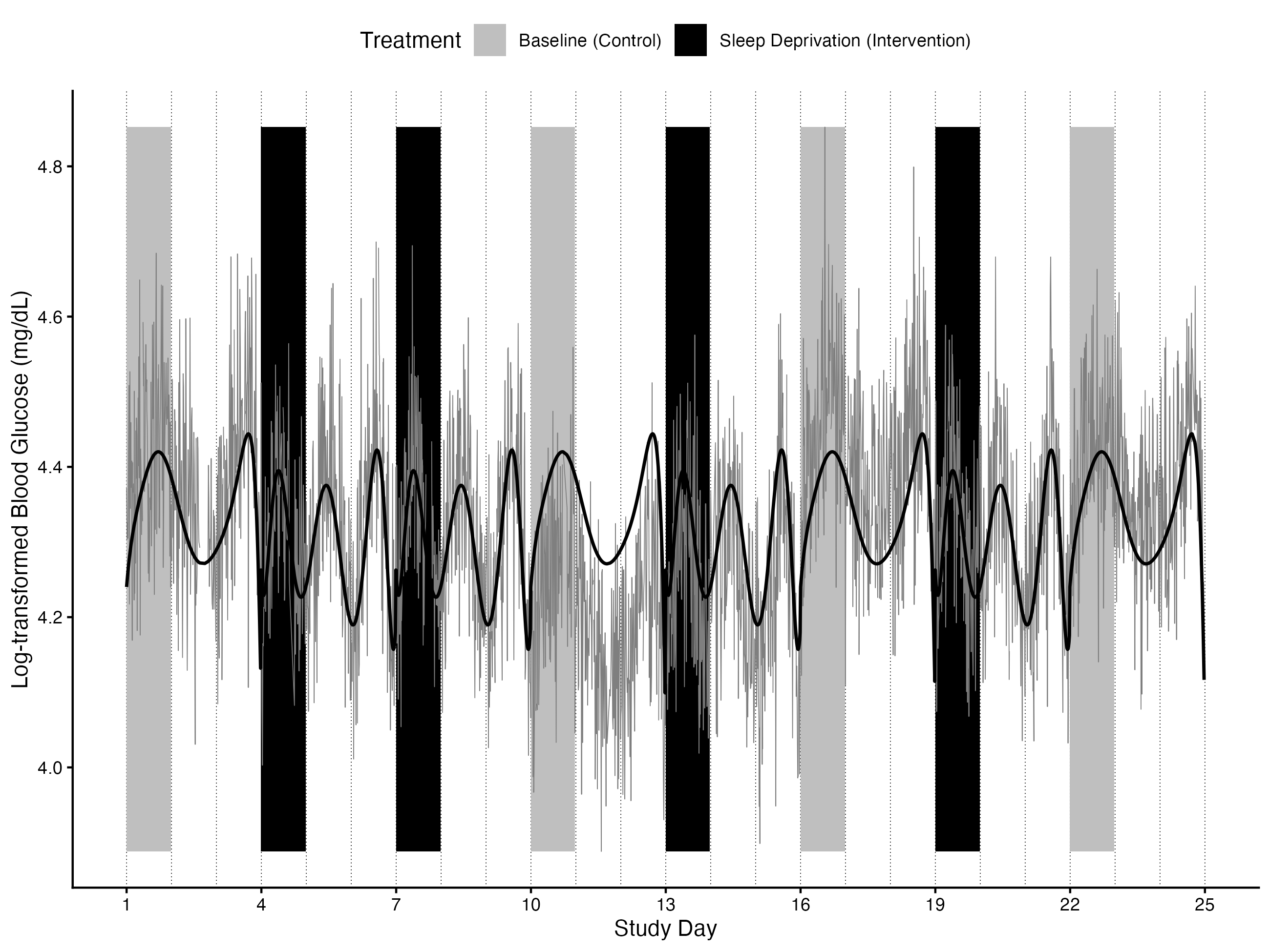}
%
%
\caption{Time series plot of simulated log-transformed blood glucose\ over m = 8 treatment periods, each with 1 treatment day (highlighted sub-period) followed by 2 washout days. Thick black line: latent mean curve estimated using a mixed-effects model in Daza et al (2019). Thin grey line: observed values simulated around mean curve. Treatment levels: baseline (grey highlight) or sleep deprivation (black highlight).}
\label{fig:time_series}       
\end{figure}

Multitudinal mixed-effects models have also been used to model esametric intensive longitudinal data, and they share deep conceptual connections to \acrshort{fosr} models \cite{fan2000two, crainiceanu2024functional}. For example, Daza et al (2019) fit a mixed-effects model over periods instead of participants.\footnote{This is an \acrshort{arco} \acrshort{glm} with period specified as a random effect.} They concluded that although ``sleep deprivation did not linearly affect log-\acrshort{bg} levels, ... [it] may have increased the {\sl variability} in latent mean log-\acrshort{bg} levels over time'' (emphasis added). This difference in variability is shown in Figure 4b of their manuscript as a {\sl pancit plot}\footnote{From Daza et al (2019): ``A pancit plot is a graph of partitioned time series segments that correspond to distinct treatment periods, and that are plotted together over the length of one period. Such plots are directly analogous to spaghetti plots, graphs of time-dependent outcomes that correspond to different study individuals in a longitudinal analysis.'' {\sl Pancit} is the Tagalog/Filipino word for ``noodles'', and is pronounced like ``pun-SEAT'' in English.} (i.e., multitudinal spaghetti plot). For didactic clarity, we have recreated this plot using ideally simulated data in Figure \ref{fig:pancit}, along with its corresponding time series plot in Figure \ref{fig:time_series}.

As a functional data analysis example, Matabuena et al (2026) \cite{matabuena2026beyond} fit a multilevel or hierarchical (i.e., multi-participant) \acrshort{fosr} model to postprandial continuous \acrshort{bg} data from four individuals, with ``day of observation'' as a period (with one meal observed on each day). They also implemented a multitudinal version of multilevel functional principal component analysis (FPCA). Figures 1 and 4 of their manuscript are pancit plots that display both the raw continuous \acrshort{bg} data and the estimated smooth average functional curves for each participant, both overall across periods and separately for each period. Figure 1(C) of Tackney et al (2026) \cite{tackney2026capturing} also displays pancit plots of electrocardiogram (ECG) data collected using a wearable sensor.

\section{Digital Clinical Science for Every One}
\label{sec:esametric_clinical_research}

We now understand how a randomized multitudinal approach can provide stronger clinical evidence---better than that of a corresponding \acrshort{rct}---for some conditions with a relatively short treatment-effect onset. For example, for conditions with rapid, reversible effects, an n-of-1 trial yields more directly relevant causal evidence for an individual who wants to understand how their treatment decisions might uniquely impact their condition based on their own health circumstances. That is, multitudinal methods are a set of core statistical approaches for reducing clinical equipoise, and therefore should be considered in designing clinical studies.

But the current infrastructure of both clinical research and clinical care is built around subgrouping approaches. This presents a substantial barrier to adopting these recurrence-based designs. Thankfully, the current regulatory and technological landscape already provides a few handholds for lifting clinical data further up the ladder of evidence.

\subsection{Bayesian Hierarchical Models and Master Protocols}
\label{subsec:master_protocol}

We previously learned how multitudinal findings can be aggregated to provide population-level clinical evidence using meta-analytic approaches or mixed-effects models. Aggregation has also been done with \acrlong{bhm}s (\acrshort{bhm}s), wherein prior distributions are assigned to group-level fixed-effects components, individual-level random-effects components, or both \cite{2010_zucker_etal, stunnenberg2015combined, senarathne2020bayesian}.

Regulatory authorities continue to recognize and support Bayesian methods as valid ways to improve clinical trial design. In particular, the recent United States \acrfull{fda} Guidance for Industry on the ``Use of Bayesian Methodology in Clinical Trials of Drug and Biological Products'' summarized how \acrshort{bhm}s have been used in \acrshort{fda} submissions to propose information-borrowing between groups \cite{us2026use}. These uses include \acrshort{bhm}s for subgroup analyses wherein each level of the hierarchy is a patient subgroup.

The \acrshort{fda} Guidance also highlighted how \acrshort{bhm}s are used to aggregate treatment effects that differ based on disease subtype. Of note: ``Bayesian analyses have been proposed for leveraging information about drug effects across related populations in {\sl basket trials} that evaluate a drug for multiple diseases or disease subtypes under a common {\sl master protocol}'' (emphases added). Now replace ``populations'' with ``individuals''. This substitution suggests a way to propose the use of a \acrshort{bhm} to aggregate multitudinal findings in a basket trial, with each patient acting as a basket.

The excerpt above more broadly suggests that multitudinal studies of single participants may be included as components or elements of more complex master protocol designs \cite{us2023master}. An adaptive design could allow each participant to add, remove, change, or continue with a treatment arm. Such a {\sl platform design} might resemble a multiphase optimization strategy (MOST) design or \acrfull{smart} \cite{collins2007multiphase, murphy2005experimental, lei2012smart}; the latter is the multi-component intervention design from which JITAIs emerged.

For example, each patient in a \acrshort{smart} multitudinal platform trial would first complete a multitudinal trial (i.e., basket) with an initial set of treatment arms. A clinical decision would then be made to modify or continue with each treatment level. The patient would then complete a subsequent multitudinal trial with the new treatment arms.

\subsection{Software Solutions}
\label{subsec:software_solutions}

We have seen how sensors enable high-resolution personalized data analysis. In addition, \acrshort{dht} platforms like software and \acrlong{app}s (\acrshort{app}s) can help patients, clinicians, and other research and care partners better coordinate the planning and implementation of multitudinal studies. Current examples include \acrshort{scd}-MVA, an \acrshort{app} for conducting \acrshort{sced}s \cite{moeyaert2021scd}, and the StudyU platform and StudyMe \acrshort{app} for conducting n-of-1 trials (respectively built for clinical researchers and non-expert individuals) \cite{Konigorski2022, zenner2022studyme}.

Esametric software and \acrshort{app}s should integrate seamlessly with the many components of a clinical study workflow in a Type 1 or Type 2 n-of-1 study design
for either clinical research or clinical care. The development of such platforms will benefit from concerted user-centered design between two co-development partners: the platform team and the clinical team (i.e., product users).

Key clinical research partners should come from organizational or ``study sponsor'' teams. These include teammates focused on science and medical aspects, biostatistics, clinical operations, safety assessment, data management, data programming, and teams that ensure patient perspectives are well-represented. Partners from interactive response technology providers and other vendors that provide logistic and technical support should also be involved.

For clinical care, important partners would include patient and caregiver representatives, clinicians, members of multidisciplinary expert panels, healthcare delivery system developers, health technology assessment professionals, hospital staff and care facility personnel, and information technology administrators.

Platform partners to involve include those from product management, data engineering, and data science teams. All co-development partners should involve their group or organization's financial decision-makers (e.g., from the business development team) early and throughout the partnership.

Other data sources of interest to platform users include sensors, medical devices, clinical records (e.g., registries, databases, laboratory testing facilities), other health \acrshort{app}s, and many other sources of personal {\sl small data} \cite{estrin2014small, hekler2019we}. To enable {\sl multimodal multitudinal analysis}, data integration should be provided as a key functionality, typically via an application programming interface (API). The platform should also support process development tracking and code/data version control.

All platform features and functionalities should be developed using coding best practices and clinically relevant good quality assurance practices (GxP)\footnote{good clinical and documentation practices (GCP, GDP), etc.}, and comply with all applicable governmental regulations. We also recommend aligning platform development decisions by using frameworks developed across multiple digital health sectors, such as the DEFINED and EVIDENCE frameworks of the Digital Medicine Society (DiMe) \cite{silberman2023rigorous, manta2021evidence}.

\subsection{The Future is Esametric}
\label{subsec:future_directions}

In a recent news release on ``Drug Repurposing to Address Unmet Medical Needs'' \cite{us20260511fdanews}, the \acrshort{fda} requested ``preliminary clinical ... [and] preclinical data'' with examples including ``data from case reports, case series, observational studies ... [and] findings from emerging tools such as artificial intelligence and machine learning''. A multitudinal study can provide better causal evidence than a case study. And n-of-1 software and \acrshort{app}s are emerging tools that can be used to create a robust evidence base for drug repurposing, decentralized trials \cite{us202409decentralized}, real-time clinical trials \cite{fda20260428rtct}, and many other modern applications.

Shifting regulatory and health governance policies will create more funding opportunities and support the business case for multitudinal designs.\footnote{Thank you to Jose Garcia for the conversation that clarified this idea.} These might include proposals for their use as a new approach methodology for in silico preclinical studies to reduce animal testing \cite{us202603nams} (for example, using digital twins \cite{qian2021synctwin, holt2024automatically}), as part of dose-finding designs for phase 1 trials, or as an individualized way to assess safety endpoints\footnote{Thank you to Richard Hahn for this idea.}.

Multitudinal approaches are increasingly being applied in precision medicine and digital health \cite{davidson2022introducing, schork2015personalized, mcdonald2021n, nikles2021creating}. Over the next few decades, advancements at the intersections of digital health, artificial intelligence and machine learning, causal inference, Bayesian approaches, and clinical trials will be particularly exciting to watch. Specific examples include n-of-1 digital phenotyping and digital biomarker development using wearable sensors \cite{shah2021personalized, vairavan2023personalized}.

The resulting innovations and breakthroughs built on decades of steady advancements in quantitative idiography will inspire more and more patients, clinicians, and researchers to adopt esametric approaches. This accelerating progress will further empower our community to pursue a better kind of health science centered on each person---both inside and outside of the clinic. With this shared vision, we will transform clinical science for every one.

\begin{acknowledgement}
My greatest thanks go to Stefan Konigorski, Suzanne McDonald, Mariola Moeyaert, Jane Nikles, Patrick Onghena, Christopher Schmid, Nicholas Schork, Linda Valeri, and other pioneering esametricians for their mentorship and guidance on multitudinal approaches over the years, and for providing early input on the structure of this chapter. I also thank my fantastic Stats-of-1 co-editors Julio Vega and Clair Robbins for their leadership and steadfast support in building our quantitative idiographic community, along with all of our excellent blog post contributors and podcast guests. Thank you to Susan Murphy for inspiring me to pursue n-of-1 trial methodology through her 2015 Greenberg Lecture at The University of North Carolina at Chapel Hill. I thank the amazing mentors and advisors whose biostatistics tutelage grounded my seminal investigations of this topic: Michael Baiocchi, Michael E. Foster, Amy Herring, Michael Hudgens, and Mark van der Laan. I thank the following people (and many others) for key insights and opportunities to spread the word about multitudinal methods in the digital health community and beyond: Richard Hahn, Eric Hekler, Jennifer Goldsack and the Digital Medicine Society, Gina Merchant, Gary Wolf and the Quantified Self; my esametric coauthor colleagues (in particular, Igor Matias, Marily Oppezzo, and Katarzyna Wac); and my many gracious hosts for multiple invited talks, seminars, interviews (in particular, Kasandra Brabaw with Fortune Magazine, Aline Holzwarth with Forbes Magazine, and the American Statistical Association), and podcast guest spots (in particular, Justin Belair, Glen Wright Colopy, Jose Garcia, Alexander Molak, and Omari Richins). Thank you to my wonderful colleagues in health technology, digital health (in particular, Luca Foschini, for his mentorship and advocacy), and the pharmaceutical industry for their camaraderie and insights. I am always so grateful for my dear family and friends; this work doesn't happen without you! Finally, I thank the editors for this rare opportunity to firmly lay the groundwork for the new field of esametry. To all who are underrepresented or unacknowledged in science, technology, engineering, and math (STEM): Kaya natin 'to! To you, the reader: Know yourself, help others, and find meaning in all things.
\end{acknowledgement}
\ethics{Competing Interests}{
The author is the founder and chief editor of Stats-of-1, a volunteer-run newsletter and podcast for science communication and advocacy. Stats-of-1 is focused on building the cross-disciplinary quantitative idiographic community that is collectively creating the modern field of esametry. The author is also a full-time employee of Boehringer Ingelheim Pharmaceuticals, Inc. The views, opinions, and statements made in this chapter are solely those of the author and may not reflect the views of Boehringer Ingelheim Pharmaceuticals, Inc. or its affiliates.
}



\section{List of Acronyms}
The following acronyms are used more than once throughout the text.
\begin{tabbing}

    \quad \quad \quad \quad \quad \= \quad \quad \quad \quad \quad \\
    
    \acrshort{app} \> \acrlong{app} \\
    \acrshort{apte} \> \acrlong{apte} \\
    \acrshort{arco} \> \acrlong{arco} \\
    \acrshort{ate} \> \acrlong{ate} \\
    \acrshort{bg} \> \acrlong{bg} \\
    \acrshort{bhm} \> \acrlong{bhm} \\
    \acrshort{cate} \> \acrlong{cate} \\
    \acrshort{cee} \> \acrlong{cee} \\
    \acrshort{dht} \> \acrlong{dht} \\
    \acrshort{fda} \> \acrlong{fda} \\
    \acrshort{fosr} \> \acrlong{fosr} \\
    \acrshort{glm} \> \acrlong{glm} \\
    \acrshort{glmm} \> \acrlong{glmm} \\
    \acrshort{hte} \> \acrlong{hte} \\
    \acrshort{ich} \> \acrlong{ich} \\
    \acrshort{ite} \> \acrlong{ite} \\
    \acrshort{jitai} \> \acrlong{jitai} \\
    \acrshort{motr} \> \acrlong{motr} \\
    \acrshort{mrt} \> \acrlong{mrt} \\
    \acrshort{pte} \> \acrlong{pte} \\
    \acrshort{rct} \> \acrlong{rct} \\
    \acrshort{rite} \> \acrlong{rite} \\
    \acrshort{scd} \> \acrlong{scd} \\
    \acrshort{sced} \> \acrlong{sced} \\
    \acrshort{smart} \> \acrlong{smart} \\

\end{tabbing}

\section{Appendix}
%
%

\subsection{More on Causal Estimands}
\label{subsec:more_on_causal_estimands}

\bigskip
\noindent
\textbf{N-of-1 Studies}
\bigskip

Let $Y_{t+1} = \mathcal{G} ( A_t, \bm{H}_t, \varepsilon_t )$ represent the structural causal mechanism (SCM) that generates the outcome at period $t+1$. In the main text, ``data-generating model'' refers to an SCM. The error term is deliberately included in the SCM expression for modeling flexibility. In statistics, this allows us to state common continuous and categorical models using the same general term; see Daza (2018) for examples. But importantly, this general expression can also denote a non-linear machine learning classification or prediction function that can be stated as a conditional expectation.

In an n-of-1 study, the main potential outcome of interest can be reasonably conceptualized as an average taken over all possible prior treatment paths. Specifically, it can be thought of as a {\sl current average potential outcome} (CAPO), defined as $Y_{t+1} ( \bullet, a_t ) = E_{\overline{A}_{t-1}} \Big\{ Y_{t+1} \big( \overline{A}_{t-1}, a_t \big) \big| A_t = a_t, \overline{\bm{V}}^\text{ex}_t \Big\}$ \cite{daza2025model}.

When treatment is randomized at every period as in an n-of-1 trial, this reduces the CAPO to $Y_{t+1} ( \bullet, a_t ) = E_{\overline{A}_{t-1}} \Big\{ Y_{t+1} \big( \overline{A}_{t-1}, a_t \big) \big| \overline{\bm{V}}^\text{ex}_t \Big\}$. Here, we consider the \acrshort{pte} equation in Section \ref{subsec:nof1_studies}, written more explicitly as:
\begin{align*}
\delta^\text{\acrshort{pte}}_{t+1}
    &= Y_{t+1} ( \bullet, 1 ) - Y_{t+1} ( \bullet, 0 ) \\
    &= E_{\overline{A}_{t-1}} \left\{
            Y_{t+1} \left( \overline{A}_{t-1}, 1 \right) -
            Y_{t+1} \left( \overline{A}_{t-1}, 0 \right)
            \Big|
            \overline{\bm{V}}^\text{ex}_t
        \right\}
\end{align*}
The resulting \acrshort{apte} across the full study period $t = 1, \dots, m$ is:
\begin{align*}
\delta^\text{\acrshort{apte}}_{(m)}
    &= E^{(m)} \left( \delta^\text{\acrshort{pte}}_{t+1} \Big| \overline{\bm{V}}^\text{ex}_m \right) \\
    &= \frac{1}{m}
        \sum_{t=1}^{m} E_{\overline{A}_{t-1}} \left\{
            Y_{t+1} \left( \overline{A}_{t-1}, 1 \right) -
            Y_{t+1} \left( \overline{A}_{t-1}, 0 \right)
            \Big|
            \overline{\bm{V}}^\text{ex}_t
        \right\}
\end{align*}
See Daza et al (2025) for the full derivations.

\bigskip
\noindent
\textbf{Single-Case Designs}
\bigskip

Earlier, we defined the \acrshort{apte} of an \acrshort{scd} by slightly modifying our earlier \acrshort{apte} definition, while adding restrictions to the set of all possible treatment vectors $\{ \overline{a}_m \}$ that describe each type of \acrshort{scd}.

For the most common types of \acrshort{scd}, the treatment vector takes the form of a {\sl phase-partitioned step-function} denoted $\overline{a}_m (\tau_1, \dots, \tau_c) = ( a_1, \dots, a_{\tau_1}$, $a_{\tau_1+1}, \dots, a_{\tau_c}$, $a_{\tau_c+1}, \dots, a_m )$, where $\tau$ indexes a crossover point, and $c$ is the number of crossover points or crossovers (i.e., the number of phases minus one).

The treatment sequences in an \acrshort{scd} are comprised of the subset of phase-partitioned step-function vectors denoted $\{ \overline{a}_m \}^\text{\acrshort{scd}} \subset \{ \overline{a}_m \}$. The most basic \acrshort{scd} with one crossover $c=1$ is simply written $\overline{a}_m (\tau) = ( a_1, \dots, a_s$, $a_{\tau+1}, \dots, a_m )$, with $\{ \overline{a}_m \}^\text{\acrshort{scd}} \equiv \overline{a}_m (\tau)$. In an AB design, the baseline phase (A) corresponds to $t = 1, \dots, \tau$, and the intervention phase (B) corresponds to $t = \tau+1, \dots, m$. The ABAB reversal design has $c=3$ crossovers with $a_t = 0$ during the baseline phases indexed by $t = 1, \dots, \tau_1$ and $t = \tau_2+1, \dots, \tau_3$, and $a_t = 1$ during the intervention phases indexed by $t = \tau_1+1, \dots, \tau_2$ and $t = \tau_3+1, \dots, m$.

In a multiple baseline design with one crossover point $c=1$, the crossover point itself $\tau_i$ is varied across the $i = 1, \dots, n$ participants; i.e., $\tau_i \in ( \tau_1, \dots, \tau_n )$. Instead of allowing only one possible step-function treatment vector as in an AB design, we would choose from a set of $n$ (not necessarily unique) step-function vectors; i.e., $\{ \overline{a}_m \}^\text{\acrshort{scd}} \equiv \{ \overline{a}_m ( \tau_1 ), \dots, \overline{a}_m ( \tau_n ) \}$.

In the presence of slow onset of the intervention effect, the \acrshort{apte} of interest might only include the subset of time points when the \acrshort{pte} is constant. That is, $\delta^\text{\acrshort{pte}}_{t'+1} = \delta^\text{\acrshort{pte}}$ over the subset indexed by $t' = 1, \dots, m'$, where $m' \le m$. This can hold, for example, if there is no autocorrelation or carryover across time points during the study. Here, effect constancy of the \acrshort{pte} implies that the \acrshort{apte} is always equal to the \acrshort{pte}; i.e., $\delta^\text{\acrshort{apte}}_{(m')} = E^{(m')} \big( \delta^\text{\acrshort{pte}}_{t'+1} \big| \overline{\bm{V}}^\text{ex}_{m'} \big) = \delta^\text{\acrshort{pte}}$.

\subsection{When are Recurring Effects Estimable?}
\label{subsec:estimable_aptes}

As with most causal quantities, for an \acrshort{apte} to be estimable, some common causal inference assumptions must hold. In our time series setting, these are the sequential versions of (causal) {\sl consistency}, {\sl ignorability}, and {\sl positivity}. One quirk of the \acrshort{apte} is that if the outcome is affected by endogenous causes, it is only {\sl partially transportable} to real-world settings where the treatment is not randomized \cite{daza2025model}. When these assumptions hold, the \acrshort{apte} can answer the multitudinal study questions shared at the start of Section \ref{subsec:bottom_up} if three other key assumptions are met.

The first assumption is that the associations between the outcome and each predictor in the SCM are {\sl stable} over time \cite{daza2018causal}. Put differently, the form of the SCM does not vary over time (i.e., $\mathcal{G}_t ( \bullet ) = \mathcal{G} ( \bullet )$ for all $t$). This is what we have assumed in our previous examples.

The second assumption is that the outcomes exhibit weak- or wide-sense stationarity (WSS) \cite{hayashi2011econometrics}. That is, $( \{ Y_{t+1} \} )$ has a constant mean and constant, finite pairwise autocorrelations. Recall that the general SCM is written as $Y_{t+1} = \mathcal{G} ( \overline{A}_t, \overline{\bm{V}}_t, \overline{Y}_t, \varepsilon_t )$, where $Y_1 = \emptyset$. Hence, for $Y_{t+1}$ to be weakly stationary, it is necessary (though not sufficient) for the covariates $\big( \big\{ \bm{V}_t \big\} \big)$ to be weakly stationary.

When the outcomes are weakly stationary, the variation in PTEs across periods is bounded and finite, and we say the \acrshort{pte} exhibits {\sl long-run effect constancy}. This implies that the \acrshort{apte} is stable or constant {\sl in the long run} for some classes of SCMs \cite{daza2018causal, daza2025model}, and the {\sl long-run \acrshort{apte}} is defined as $\lim_{m \rightarrow \infty} \delta^\text{\acrshort{apte}}_{(m)} = \delta^\text{\acrshort{apte}}$. While this long-run \acrshort{apte} is a recurring quantity, it may not be identifiable.

The third assumption is that the number of endogenous SCM predictors (i.e., elements of $\overline{\bm{V}}_t$ and $\overline{Y}_t$) is finite and small enough to be estimated with a realistically sized dataset. This is called the {\sl finite endogeneity} assumption \cite{daza2025model}. For example, suppose the SCM is $Y_{t+1} = \mathcal{G} \big( \overline{A}_t, \overline{Y}^\ell_t, \varepsilon_t \big)$, where $\overline{Y}^\ell_t = ( Y_{t-\ell}, \hdots, Y_{t-2}, Y_{t-1}, Y_t )$ denotes the inclusive history vector (i.e., set of all previous values of $Y_t$ and $Y_t$ itself) up to lag $\ell = 0, 1, \hdots, t-1$. If $\ell = 2$, then $\overline{Y}^2_t = ( Y_{t-2}, Y_{t-1}, Y_t )$ may be small enough to allow us to collect enough data to fit an outcome model. The second and third assumptions together have been called the {\sl finite time horizon} assumption \cite{daza2025model}.

An \acrshort{apte} that meets all of these assumptions is a viable \acrshort{rite} estimand for a multitudinal study. It is therefore important to test these assumptions when possible (some are not testable). For example, sequential positivity can be checked by examining pairwise cross-correlations between the predictors $\big\{ \overline{A}_t, \overline{\bm{V}}_t, \overline{Y}_t \big\}$. To test the WSS assumption, we can use the Augmented Dickey Fuller (ADF) and Kwiatkowski-Phillips-Schmidt-Shin (KPSS) unit-root tests for stationarity \cite{daza2018causal}.\footnote{If a continuous outcome is not stationary, it may be tempting to apply a ``stationarization'' method to the outcome, like pre-whitening (i.e., taking first differences) or de-trending. If successful, the transformed outcome variable is stationary. However, transforming the outcome may require changing the study's original scientific or clinical hypothesis. \cite{daza2018causal}}

\subsection{Multitudinal Generalized Linear Mixed Model}
\label{subsec:arcoglmm}

We can use the \acrshort{arco} model terminology to define a multitudinal \acrshort{glmm} as $g \big( \mu_{t+1} ( \bm{Z}_t ) \big) = \bm{\eta}_t + \bm{\zeta}_t$, where:
\begin{itemize}

    \item $\mu_{t+1} ( \bm{Z}_t ) = E \big( Y_{t+1} \big| A_t, \overline{A}_{t-1}, \overline{Y}_t, \bm{V}^\text{ex}_t, \bm{Z}_t \big)$ is the mean outcome conditional on the treatment $A_t$, all other model predictors, and random-effect indicators $\bm{Z}_t$

    \item $\bm{\eta}_t$ is the fixed-effects linear predictor

    \item $\bm{\zeta}_t$ is the random-effects linear predictor

    \item $g ( \bullet )$ is the \acrshort{glmm} link function
    
\end{itemize}

Here, the fixed-effects linear predictor is
\begin{align*}
    \bm{\eta}_t
        & = \beta_0 +
            \beta_\text{A} A_t +
            \overline{A}^{\ell_A}_{t-1} \bm{\beta}_\text{co} +
            \bm{A}^\otimes_t \bm{\beta}_\text{Aco} +
            \overline{Y}^{\ell_Y}_t \bm{\beta}_\text{ar} +
            \bm{Y}^\otimes_t \bm{\beta}_\text{Aar} +
            \bm{V}^\text{ex}_t \bm{\beta}_\text{ex}
.
\end{align*}
The coefficient vector $\bm{A}^\otimes_t$ denotes all unique two-way interactions between $A_t$ and the elements of $\overline{A}^{\ell_A}_{t-1}$. Likewise, $\bm{Y}^\otimes_t$ is the vector of all unique two-way interactions between $A_t$ and the elements of $\overline{Y}^{\ell_Y}_t$. The $\beta$ terms denote fixed-effect coefficients as defined in the literature on mixed-effects models. Similarly, the random-effects linear predictor is
\begin{align*}
    \bm{\zeta}_t
        & = u_0 +
            u_\text{A} A_t +
            \overline{A}^{\ell_A}_{t-1} \bm{u}_\text{co} +
            \bm{A}^\otimes_t \bm{u}_\text{Aco} +
            \overline{Y}^{\ell_Y}_t \bm{u}_\text{ar} +
            \bm{Y}^\otimes_t \bm{u}_\text{Aar} +
            \bm{V}^\text{ex}_t \bm{u}_\text{ex}
,
\end{align*}
where the $u$ terms denote random-effect coefficients.

\backmatter
\printindex


\end{document}